\documentclass[aip,jmp,reprint,amsmath,amssymb]{revtex4-1}

\usepackage{graphicx}
\usepackage{dcolumn}
\usepackage{bm}
\usepackage{color}

\begin{document}

\title{Matter wave switching in Bose-Einstein condensates via intensity redistribution soliton interactions}

\author{S.~Rajendran}
\affiliation{Centre for Nonlinear Dynamics, Bharathidasan University, Tiruchirapalli 620024, India}

\author{P. Muruganandam}
\affiliation{Department of Physics, Bharathidasan University, Tiruchirapalli 620024, India}

\author{M. Lakshmanan}
\affiliation{Centre for Nonlinear Dynamics, Bharathidasan University, Tiruchirapalli 620024, India}

\begin{abstract}

Using time dependent nonlinear (\emph{s}-wave scattering length) coupling  between the components of a weakly interacting two component Bose-Einstein condensate (BEC), we show the possibility of matter wave switching (fraction of atoms transfer) between the components via shape changing/intensity redistribution (matter redistribution) soliton interactions. We investigate the exact bright-bright $N$-soliton solution of an effective one-dimensional (1D) two component BEC by suitably tailoring the trap potential, atomic scattering length and atom gain or loss. In particular, we show that the effective 1D coupled Gross-Pitaevskii (GP) equations with time dependent parameters can be transformed into the well known completely integrable Manakov model described by coupled nonlinear Schr\"odinger (CNLS) equations by effecting a change of variables of the coordinates and the wave functions under certain conditions related to the time dependent parameters.  We obtain the one-soliton solution and demonstrate the shape changing/matter redistribution interactions of two and three soliton solutions for the time independent expulsive harmonic trap potential, periodically modulated harmonic trap potential and kink-like modulated harmonic trap potential. The standard elastic collision of solitons occur only for a specific choice of soliton parameters.

\end{abstract}


\maketitle

\section{Introduction}

The past decade has witnessed a considerable increase  of interest in the  experimental and theoretical studies of matter wave solitons of the dark~\cite{Burger1999, Denschlag2000,Frantzeskakis2010} and bright~\cite{Strecker2002, Khaykovich2002} types in Bose-Einstein Condensates (BECs). These solitons  have attracted a great deal of attention in connection with the dynamics of nonlinear matter waves, including soliton propagation \cite{Busch2000, Salasnich2004}, vortex dynamics~\cite{Rosenbusch2002}, interference patterns~\cite{Liu2000} and domain walls in binary  BECs~\cite{Trippenbach2000, Malomed2004}.  From an experimental BEC point of view, bright solitons are created themselves as condensates~\cite{Strecker2002, Khaykovich2002} while dark solitons exist as notches or holes within the condensates~\cite{Burger1999, Denschlag2000}. Note that bright solitons propagate over much larger distances than dark solitons. The bright matter-wave soliton trains (multi-solitons) were experimentally observed by Strecker et al.~\cite{Strecker2002} in $^7$Li and Khaykovich et al.~\cite{Khaykovich2002} in $^{87}$Rb by magnetically tuning the atomic scattering length from repulsive to attractive nature, through Feshbach resonance~\cite{Chin2010}. The recent experiments at Heidelberg and Hamburg universities have shown the formation of dark solitons, their oscillations and interaction in single component BECs of $^{87}$ Rb atoms with confining harmonic potential~\cite{Christoph2008, Weller2008, Stellmer2008}.

Many aspects of the above novel and experimentally accessible form of matter have been since then intensively studied; one of them concerns with the investigation of the behaviour of multi-component BECs, which have been experimentally studied in either mixtures of different hyperfine states of the same atomic species or even in mixtures of different atomic species. Experimental generation of two-component BECs of different hyperfine states of rubidium atoms in a magnetic trap~\cite{Myatt1997} and of sodium atoms in an optical trap~\cite{stamper98} stimulated theoretical studies devoted to the mean-field dynamics of multi-component condensates~\cite{Wadati2004}. Recently, Zhang et al.~\cite{Zhang2009} proposed a method for independent tuning of scattering lengths in multi-component BECs.  When a condensate is cigar shaped and has relatively low density, that is, when the healing length of  the components is much larger than the transverse dimension of the condensate and much less than its longitudinal dimension, the transverse atomic distribution is well approximated  by the Gaussian ground state and the system of coupled 1D  Gross-Pitaevskii (GP) equations is adequate to describe multi-component condensates.

 In the recent literature, there has been a growing interest, both from experimental as well as theoretical perspectives~\cite{Anglin1997, Ruostekoski1998, Vardi2001, Ponomarev2006, Wang2007, Rajendran2009}, in studying the dynamics of two-component BECs coupled to the environment such as external thermal clouds which leads to the mechanism of loading (gain) external atoms (thermal clouds) into the BECs by optical pumping or continuously depleting (loss of) atoms. It is interesting to note that matter wave solitons in multi-component BECs hold promise for a number of applications, including the multi-channel signals and their switching, coherent storage and processing of optical fields~\cite{Folman2000,Folman2001,Petrov2009}.

Further, there has been increased interest in recent times in studying the properties of BECs with time varying control parameters, such as (i) the temporal variation of atomic scattering length which can be achieved through Feshbach resonance~\cite{Moerdijk1995,Roberts1998,Stenger1999,Cornish2000}, (ii) inclusion of appropriate time dependent gain or loss terms~\cite{Anglin1997, Ruostekoski1998, Syassen2008, Kneer1998, Davis2000, Drummond2000, kohl2002, Vardi2001, Ponomarev2006, Wang2007, Rajendran2009}, (iii) the temporal modulation of trap frequencies~\cite{Rajendran2010, Rajendran2009, Janis2005, Baizakov2005,Mayteevarunyoo2009, Atre2006} and so on. In particular, the study of matter wave solitons under time varying control parameters is one of the current active research fields~\cite{Atre2006, Serkin2007, Li2008, Zhao2008, Zhao2009, Zhao2009a, Rajendran2009, Serkin2010, Rajendran2010}. Being motivated by the above considerations, in this paper, we study the dynamics of the exact bright-bright matter wave solitons of two component BECs under the above time varying control parameters.

Most of the theoretical studies on the matter wave solitons in multi-component BECs have been carried out using either numerical or approximation methods. For example, two component dark-bright solitons have been reported in~\cite{Busch2001}; bound dark solitons have been numerically studied in~\cite{ohberg01}, where it has been found that the creation of slowly moving objects is possible; a diversity of other bound states has been generated numerically in~\cite{Kevrekidis2005}. The present authors have obtained the exact dark-bright soliton solutions and their different kinds of interactions of the two component BECs by suitably tailoring the trap potential, atomic scattering length and atom gain/loss~\cite{Rajendran2009}.  We also point out that Babarro~et~al.~\cite{Babarro2005} have shown the possibility of switching phenomenon of matter-wave solitons via bright-bright soliton interaction of different species of two component BECs essentially without potential term or any modulation of control parameters, through an approximate theoretical analysis and numerical simulations. In the present paper, we have investigated a different kind of matter wave switching phenomenon (in two different hyperfine states of the same species such as $^{85}$Rb with equal intra- and inter-species atomic interactions) via intensity redistribution of \emph{exact} bright-bright soliton interactions with the temporally modulated control parameters analytically.

Specifically, in the present paper we bring out the exact bright-bright one-, two-, three- and $N$-soliton solutions in two-component BECs by simply mapping a class of two coupled effective 1D GP equations onto the completely integrable two coupled nonlinear Schr\"odinger (2CNLS) equations (Manakov system) and making use of the exact $N$-soliton solutions of the latter system. In particular we have demonstrated the bright-bright shape changing/matter redistribution of two and three solitons under collision, while elastic collision occurs for a very special choice parameters. We have also shown the shape restoring property in the case of three soliton interaction. These types of elastic and shape changing interactions of two and three solitons have been well studied in the context of optical computing, where the intensity of light pulses are transformable between two modes of Manakov type systems~\cite{Radhakrishnan1997, Steiglitz2000, kanna01, kanna03, Vijayajayanthi2009}. From the BEC point of view the shape changing interaction can be interpreted as the transformation of the fraction of atoms between the components, which is the so called  matter wave switch. Such matter wave switching phenomenon can be used to manipulate matter wave devices such as switches, logic gates and atom-chip~\cite{Folman2000, Folman2001, Petrov2009}. One of the long term prospectives of matter wave devices is their potential application to quantum information processing, for details see for example Refs.~\cite{Folman2000, Folman2001, Petrov2009}. In the present paper we have shown that such a matter wave switching phenomenon in two component BECs is possible via shape changing soliton interactions by suitably tailoring the trap potential, atomic scattering length and gain or loss term.

This paper is organized as follows. In Section~\ref{sec.Ansatz}, we present the ansatz for the two coupled GP equations in 1D to be mapped onto the integrable 2CNLS equations. In Section ~\ref{sec.Multi-Soliton solutions} we deduce the one-, two-, three- and $N$-soliton solutions of the two coupled GP equation from the one-, two-, three- and $N$-soliton solutions, respectively, of the 2CNLS equations. In section~\ref{sec.Elastic and shape changing interactions of bright-bright solitons}, we bring out the one soliton solution, elastic and shape changing soliton interaction of two and three soliton solutions for different forms of the trap potential, gain/loss term and interatomic interaction (scattering length). We have also shown the shape restoring property in three soliton interactions. Elastic collision occurs only for a specific choice of parameters. The analysis can also be extented to $N$-solitons without much difficulty. Finally, in Section~\ref{sec.Conclusion}, we present a brief summary of our study.

\section{Ansatz for mapping two Coupled GP equations onto Manakov system}\label{sec.Ansatz}

We consider the dynamics of a 1D  two-component  trapped BEC with gain or loss term  by the mean-field equations for the wave functions, $\Psi_1$ and $\Psi_2$, of the condensates $\vert 1\rangle$ and $\vert 2\rangle$:
\begin{align}
i\frac{\partial \Psi_{j}}{\partial t} =  &\, -\frac{1}{2}\frac{\partial^2 \Psi_{j}}{\partial  x^2}  + \Biggl[ R(t) \sum_{k=1}^{2} \sigma_{jk}\vert \Psi_{k}\vert^2  \notag \\
&\, + V(x,t) -\mu_j+\frac{i}{2} \gamma(t) \Biggr] \Psi_{j}, \;\;j = 1,2, \label{cgpe}
\end{align}
 where $V(x,t)= \Omega^2(t) x^2/2$ is the external time varying trap potential, which is expulsive for $\Omega^2(t) < 0$ and confining for $\Omega^2(t) > 0$. Here $\Omega^2(t)=\omega_x^2(t)/\omega_{\perp}^2$, $\omega_x(t)$ is the temporally modulated axial trap frequency, $\omega_{\perp}$ is the time independent radial trap frequency, $R(t)=2 a_s(t)/a_B$, $a_s(t)$ is the magnitude of the $s$-wave atomic scattering length, $a_B$ is the Bohr radius, $\sigma_{jk}$'s are the signs of the $s$-wave scattering  lengths, which are negative for attractive and positive for repulsive interactions, $\mu_j$ is the chemical potential of the $j$th component and $\gamma(t)=\Gamma(t)/\omega_{\perp}$, where $\Gamma(t)$ is the gain (for $+$ve) or loss (for $-$ve) term, which is the phenomenologically incorporated interaction of external atoms (thermal clouds). The time dependent gain/loss term corresponds to the mechanism of continuously loading external atoms into the BEC (gain) by optical pumping or continuous depletion (loss) of atoms from the BEC~\cite{Syassen2008,Kneer1998,Davis2000,Drummond2000,kohl2002}.

The atomic scattering length of alkali atoms such as $^7$Li and $^{85}$Rb atoms can be experimentally varied by suitably tuning the external magnetic field through the Feshbach resonance as~\cite{Strecker2002, Khaykovich2002, Courteille1998}
\begin{align}\label{feshbach}
a_s(t)=a_s^0\left(1-\frac{\Delta}{B(t)-B_0}\right),
\end{align}
where $a_s^0$ is the scattering length of the condensed atoms, $B(t)$ is the external time varying magnetic field, $B_0$ is the resonance magnetic field and $\Delta$ is the resonance width. Recently, Zhang et al.~\cite{Zhang2009} proposed a method for a similar kind tuning of scattering lengths for two component condensates of two different hyperfine states of $^{87}$Rb. In the present study we have considered the attractive-attractive ($\sigma_{jk}= -1$) two component BECs as in the case of two different hyperfine states of $^{85}$Rb with equal intra- and inter-species atomic interactions.

By applying the following transformation,
\begin{align}
\Psi_j(x,t)= \exp\left[i \mu_j t+\int \frac{\gamma(t)}{2} dt\right] Q_j(x,t), \label{trans}
\end{align}
where $j=1,2$, Eq.~(\ref{cgpe}) can be transformed to a system of two coupled GP equations without gain/loss  term and chemical potential as
\begin{align}
i\frac{\partial Q_j}{\partial t} =  &\,  \left[\tilde{R}(t) \sum_{k=1}^{2} \sigma_{jk}\vert Q_k\vert^2  + V(x,t) -\frac{1}{2}\frac{\partial^2 }{\partial  x^2}\right] Q_j, \label{cgpe1}
\end{align}
where $j = 1,2$ and
\begin{align}
\tilde{R}(t)= \exp \left[ \int \gamma(t) dt \right] R(t) \label{rtilde}.
\end{align}
Eq.~(\ref{cgpe1}) can be mapped onto the 2CNLS (Manakov) equations under the following transformation~\cite{Gurses2007, Serkin2007, Kundu2009, Rajendran2010, Serkin2010},
\begin{align}
Q_j(x,t) = \Lambda q_j(X,T),  \quad j=1,2 , \label{qtrs}
\end{align}
where the new independent variables $T$ and $X$ are chosen as functions of the old independent variables $t$ and $x$ as
\begin{align}
T=&G(t), \;\;\; X=F(x,t), \label{FG}
\end{align}
while $\Lambda = \Lambda(x,t)$ is a function of $t$ and $x$. Applying the above transformation (\ref{qtrs}), so that
\begin{align}
\frac{\partial}{\partial t} &= G_t \frac{\partial}{\partial T}+ F_t \frac{\partial}{\partial X},  \qquad \frac{\partial}{\partial x}= F_x \frac{\partial}{\partial X}, \notag \\ \frac{\partial^2}{\partial x^2} &= F_{xx} \frac{\partial}{\partial X}+ F_x^2 \frac{\partial^2}{\partial X^2}, \quad (G_t= \frac{\partial G}{\partial t}, F_x= \frac{\partial F}{\partial x}),
\end{align}
one can reduce Eq.~(\ref{cgpe1}) to the 2CNLS equation of the form
\begin{align}
i\frac{\partial q_j}{\partial T}+ \frac{\partial^2 q_j}{\partial  X^2} + 2 \left(\sum_{k=1}^{2}  \vert q_k\vert^2\right) q_j=0, \;\; j = 1, 2, \label{2cnls}
\end{align}
subject to the conditions that the functions $\Lambda$, $F$, $G$, $\Omega$ and
$\tilde{R}$ should satisfy the following set of equations,
\begin{subequations}
\begin{align}
& i \Lambda_t + \frac{1}{2}\Lambda_{xx}-\frac{\Omega(t)}{2} x^2 \Lambda=0, \label{eq:cond:a}\\
& i \Lambda F_t + \frac{1}{2}\left(2 \Lambda_{x} F_x+ \Lambda F_{xx}\right)=0, \\
& G_t = \frac{F_x^2}{2} = \frac{\tilde{R}}{2} \vert \Lambda \vert ^2. \label{eq:cond:c}
\end{align}\label{eq:cond}
\end{subequations}
In order to solve for the unknown functions $\Lambda$, $F$ and $G$ in the above equations (\ref{eq:cond}) we assume the polar form%
\begin{align}
\Lambda=r(x,t) \exp[i \theta(x,t)].\label{Lambda}
\end{align}
One can immediately check from the relations~(\ref{eq:cond:c}) that $r$ is a function of $t$ only, $r=r(t)$, since $G$ and $\tilde{R}$ are functions of $t$ only. Then from Eqs.~(\ref{eq:cond}) one can easily deduce the transformation function $\Lambda$ given by~(\ref{Lambda})  and the transformations $G(t)$ and $F(z,t)$ through the following relations,
\begin{subequations} \label{x-and-t}
\begin{align}
r^2  & = 2 r_0^2  \tilde{R},\\
\theta & = -\frac{\tilde{R}_t}{2 \tilde{R}} x^2 +  2 b r_0 ^2 \tilde{R} x- 2 b^2 r_0^4 \int \tilde{R}^2 dt, \\
F(x,t) &= X = \sqrt{2} r_0 \tilde{R} x-2 \sqrt{2} b r_0^3 \int \tilde{R}^2 dt, \\
G(t) &=T = r_0^2  \int \tilde{R}^2 dt.
\end{align}
\end{subequations}
Here $b$, $r_0$ are arbitrary constants, and $\tilde{R}$ and $\Omega^2$ should be related by the following condition,
\begin{align}
\frac{d}{dt}\left(\frac{\tilde{R}_t}{\tilde{R}}\right) -\left(\frac{\tilde{R}_t}{\tilde{R}}\right)^2-\Omega^2(t)=0, \label{riccati}
\end{align}
which is a Riccati type equation for $\tilde{R}_t/\tilde{R}$. Eq.~(\ref{2cnls}) is the celebrated Manakov system~\cite{Agrawal1995, Ablowitz2000, Hasegawa2000, Scott1984}, which is well studied in the context of nonlinear optics, biophysics, plasma physics etc. Eq.~(\ref{2cnls}) is a completely integrable soliton system and exhibits interesting one-, two-, three- and $N$-soliton solutions of bright-bright type~\cite{Radhakrishnan1997, kanna01, kanna03, Vijayajayanthi2009}. From the solutions of Eq.~(\ref{2cnls}), one can straightforwardly construct the one-, two-, three- and $N$-bright-bright soliton solutions for Eq.~(\ref{cgpe}), provided ${R}_t$, $\gamma(t)$ and $\Omega(t)$ satisfy Eq.~(\ref{riccati}). In the context of BECs, it is of fundamental interest to study the bright-bright soliton solutions of Eq.~(\ref{cgpe}). In the following we shall describe the  multi-soliton solutions corresponding to the coupled GP equation (\ref{cgpe}).

Note that in Eq.~(\ref{cgpe}), the variable $x$ actually represents $x/a_{\perp}$, where $a_{\perp}=\hbar / (m \omega_{\perp})$, and similarly $t$ stands for $\omega_{\perp} t$.

\section{Multi-Soliton solutions of two coupled nonlinear Schr\"odinger equations}\label{sec.Multi-Soliton solutions}

The 2CNLS equation~(\ref{2cnls}) in contrast to the single component NLS system admits solutions which exhibit certain novel energy sharing (shape changing) collisions~\cite{Radhakrishnan1997, kanna01, kanna03, Vijayajayanthi2009}. Recently, the general expression for  $N$-soliton solution of the Manakov system in the Gram determinant form has been given in Ref.~\cite{Vijayajayanthi2009} by using Hirota's bilinear method.

In order to write down the multi-soliton ($N$-soliton) solution of the focusing Manakov system~(\ref{2cnls}), we define the following $(1\times N)$ row matrix $C_s$, $s=1,2$, $(2\times 1)$ column matrices $\psi_j$, and $(N\times 1)$ column matrix $\phi$, $j=1,2,\ldots, N$, and the $(N\times N)$ identity matrix $I$:
\begin{subequations}
\begin{align}
C_s= -\left(\alpha_1^{(s)}, \alpha_2^{(s)}, \ldots, \alpha_{N}^{(s)}\right),\quad  {\bf{0}}=(0, 0, \ldots, 0),\quad \quad	
\end{align}
\begin{align}
\psi_j=
\begin{pmatrix}
\alpha_j^{(1)}\\
\alpha_j^{(2)}
\end{pmatrix},
\phi=
\begin{pmatrix}
e^{\eta_1}\\
e^{\eta_2}\\
\vdots\\
e^{\eta_{N}}
\end{pmatrix},
I=
\begin{pmatrix}
1 & 0 & \cdots & 0\\
0 & 1 & \cdots & 0\\
\vdots & \vdots & \ddots &\vdots\\
0 & 0 & \cdots &1
\end{pmatrix}.
\end{align}
\end{subequations}
Here $\alpha_{j}^{(s)}$, $s=1,2$, $j=1,2,\ldots, N$, are arbitrary complex parameters and $\eta_i=k_iX+ik_i^2T$, $i=1,2, \ldots, N$, and $k_i$ are
complex parameters.  We write down the multi-soliton solution of the 2-CNLS system~\cite{Vijayajayanthi2009} as below:
\begin{align}
q_s(X,T) = \frac{g^{(s)}}{f}, \;\;\; s=1, 2, \label{sol:hirota}
\end{align}
where
\begin{subequations}
\label{sol:manko}
\begin{align}
g^{(s)}=
\left\vert
\begin{matrix}
A & I & \phi\\
-I & B^T & {\bf 0}^T\\
{\bf 0} & C_s & 0
\end{matrix}
\right\vert, \;\;\;\; f= \left\vert
\begin{matrix}
A & I\\
-I & B^T
\end{matrix}
\right\vert.\label{f}
\end{align}
Here the matrices $A$ and $B$ are defined as
\begin{align}
A_{ij}= \frac{e^{\eta_i+\eta_j^*}}{k_i+k_j^*}, \; B_{ij}=\frac{\psi_j^{\dagger}\psi_i}{k_j^*+k_i},\; i,j=1, 2, \ldots, N.  \label{omg}
\end{align}
\end{subequations}%
In equation (\ref{omg}), $\dagger$ represents the transpose conjugate.
In particular the one-soliton solution ($N=1$ case) reads as
\begin{align}
q_j(X,T)=\frac{\left\vert
\begin{matrix}
A_{11} & 1 & e^{\eta_1}\\
-1 & B_{11} & 0\\
0 & -\alpha_{1}^{(j)} & 0
\end{matrix}\right\vert}{\left\vert
\begin{matrix}
A_{11} & 1\\
-1 & B_{11}
\end{matrix}
\right\vert}=\frac{\alpha_{1}^{(j)} e^{\eta_1}}{1+\beta_0 e^{\eta_1+\eta_1^*}},
\end{align}
where $j=1,2 $ and
\begin{align}
\beta_0=\frac{\vert \alpha_{1}^{(1)}\vert^2+\vert \alpha_{1}^{(2)}\vert^2}{(k_1+k_1^*)^2}
\end{align}
For the $N=2$ case, we can write  the two-soliton solution  from Eq.~(\ref{sol:hirota}) as
\begin{align}
q_j({X,T})=\frac{\left\vert
\begin{matrix}
A_{11} & A_{12} & 1 & 0& e^{\eta_1}\\
A_{21} & A_{22} & 0 & 1& e^{\eta_2}\\
-1 & 0&B_{11}& B_{21}& 0\\
0 & -1&B_{12}& B_{22}& 0\\
0 & 0 & -\alpha_{1}^{(j)}& -\alpha_{2}^{(j)} & 0
\end{matrix}\right\vert}{\left\vert
\begin{matrix}
A_{11} &  A_{12} & 1 & 0\\
A_{21} & A_{22} & 0 & 1\\
-1 & 0&B_{11}& B_{21}\\
0 & -1&B_{12}& B_{22}
\end{matrix}
\right\vert},
\end{align}
with $j=1,2$. The explicit form of the two-soliton solution can be written as
\begin{align}
q_j({X,T})= \frac{1}{D} \sum_{l=1}^{2} \displaystyle\left(\alpha_{l}^{(j)} e^{\eta_l}+ \xi_l^{(j)} e^{\eta_1+\eta_2+\eta_l^*}\right), \label{eqn.20}
\end{align}
where
\begin{align}
&D=1+\sum_{l,m=1}^{2}\displaystyle \left(\beta_{lm} e^{\eta_l+\eta_m^*}\right)+\displaystyle \beta_{3} e^{\eta_1+\eta_l^*+\eta_2+\eta_2^*}\notag \\
&\xi_l^{(j)}=\displaystyle \frac{k_1-k_2}{(k_1+k_l^*)(k_2+k_l^*)} \left(\alpha_1^{(j)} \kappa_{2l}-\alpha_2^{(j)}\kappa_{1l}\right), \,  \notag\\
&\beta_3 =\frac{\vert k_1-k_2 \vert^2  (\kappa_{11} \kappa_{22}-\kappa_{12}\kappa_{21})}{(k_2+k_2^*)(k_2+k_1^*) \vert k_1+k_2^* \vert^2}, \notag \\
 &\beta_{lm}=\frac{\kappa_{lm}}{k_l+k_m^*}, \quad\kappa_{lm}=\sum_{j=1}^2 \dfrac{ \alpha_l^{(j)} \alpha_m^{(j)*}}{k_l+k_m^*} \label{eqn.21}.
\end{align}
Similarly, the three-soliton and $N$-soliton solutions can be written down explicitly.

Now using the transformation (\ref{trans}), the bright-bright $N$-soliton solution of the two-coupled GP equations can be written as
\begin{align}
\Psi_j(x,t)=&2 r_0\sqrt{ \tilde R} \exp\left[ i (\theta+\mu_j t)+\int \frac{\gamma}{2} dt\right] q_j(X,T), \label{sol:cgpe}
\end{align}
where $X$ and $T$ are given in Eqs.~(\ref{x-and-t}c) and (\ref{x-and-t}d).
 Note that the variables $X$ and $T$ are complicated functions of the original  independent variables $x$ and $t$ as given by Eqs.~(12c) and (12d). Consequently the width which is inversely proportional to $\tilde{R}(t)$, see Eq. (12 c), changes as a function $t$ and $x$, even though it is not so in the case of the Manakov system.

\section{Elastic and intensity redistribution interactions of BEC bright-bright solitons} \label{sec.Elastic and shape changing interactions of bright-bright solitons}

Depending on the various forms of the trap potential, gain/loss and interatomic interaction satisfying Eq.~(\ref{riccati}), we have deduced different kinds of bright-bright soliton solutions using the above expression for the soliton solutions (\ref{sol:cgpe}), and the transformations (\ref{qtrs}), (\ref{Lambda}) and (\ref{x-and-t}). In the following, we demonstrate soliton solutions for three typical choices of trap potentials. In particular, we have focussed on the  shape changing  and elastic interactions of two soliton ($N=2$) and three-soliton ($N=3$) solutions of the two component BEC. The analysis can be systematically extended to arbitrary $N$, which we do not present here for brevity. 

\subsection{Expulsive Potential}

 For time independent expulsive parabolic trap potential, $\Omega^2(t)= -\Omega_0^2$, where $\Omega_0$ is a constant, we get $\tilde{R}(t) = \mbox{sech} (\Omega_0 t + \delta)$ from Eq.~(\ref{riccati}). The intensity of the wave packet ($\vert \Psi_j \vert ^2$) is proportional to $ \tilde{R}(t) e^{\int \gamma(t) dt}$ as seen from Eq.~(\ref{sol:cgpe}) and the width of the wave packet is inversely proportional to $ \sqrt{\tilde{R}(t)}$. We have constructed different types of soliton solutions by suitably tuning the gain term but here we have presented only the constant intensity case for $\gamma=\Omega_0 \tanh(\Omega_0 t + \delta)$.
\begin{figure}[!ht]
\begin{center}
\includegraphics[width=0.9\linewidth,clip]{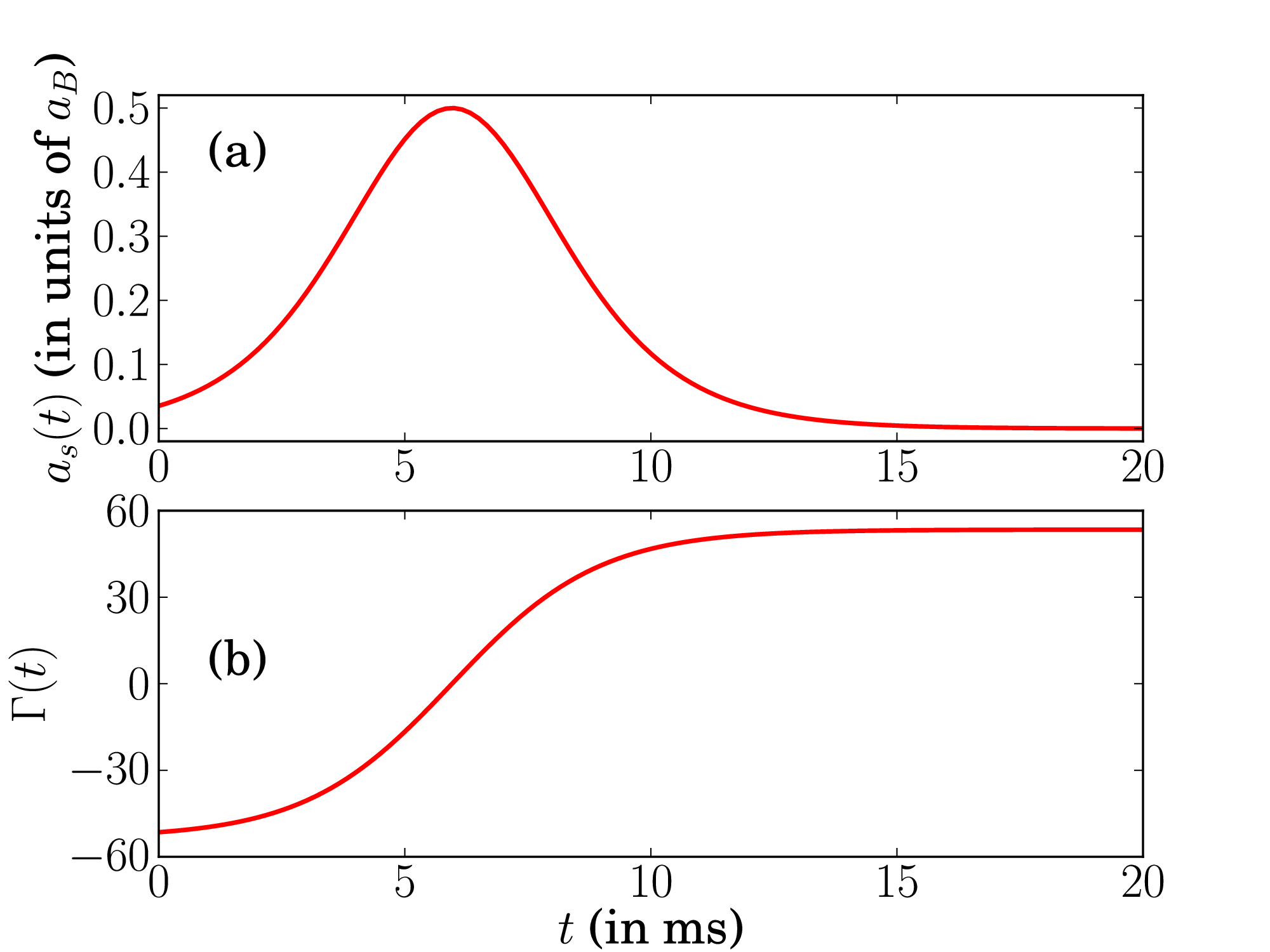}
\caption{(color online) Choice of the atomic scattering length $a_s(t)$ and gain term $\Gamma(t)$ as a function of time for time-independent expulsive harmonic trap potential $V(x) =-\Omega_0^2 x^2/2$, $\Omega_0 = \text{constant}$.}
\label{fig_gamma_1}
\end{center}
\end{figure} 
Fig.~\ref{fig_gamma_1} shows one such possible gain term $\Gamma(t) = \omega_\perp \Omega_0 \tanh(\Omega_0 t + \delta)$ and  the atomic scattering length $a_s(t) = \frac{1}{2} a_BR(t) = \frac{1}{2}a_B\, \mbox{sech}^2(\Omega_0 t + \delta)$, which can be experimentally realized by varying the external magnetic field as
\begin{align}
B(t)=B_0+ \frac{ a_s^0 \,\,\, \Delta}{a_s^0 - \frac{1}{2} a_B \mbox{sech}^2(\Omega_0 t+\delta)}.
\end{align}%
One may note that such a form of scattering length $a_s(t)$ has been realized in $^{7}$Li and $^{85}$Rb atoms~\cite{Strecker2002, Khaykovich2002, Courteille1998}. Similarly the gain term $\Gamma(t)$ can be experimentally realized by pumping of atoms optically as demonstrated in Refs.~\cite{Davis2000,kohl2002}.
\begin{figure}[!ht] 
\begin{center}
\includegraphics[width=\linewidth,clip]{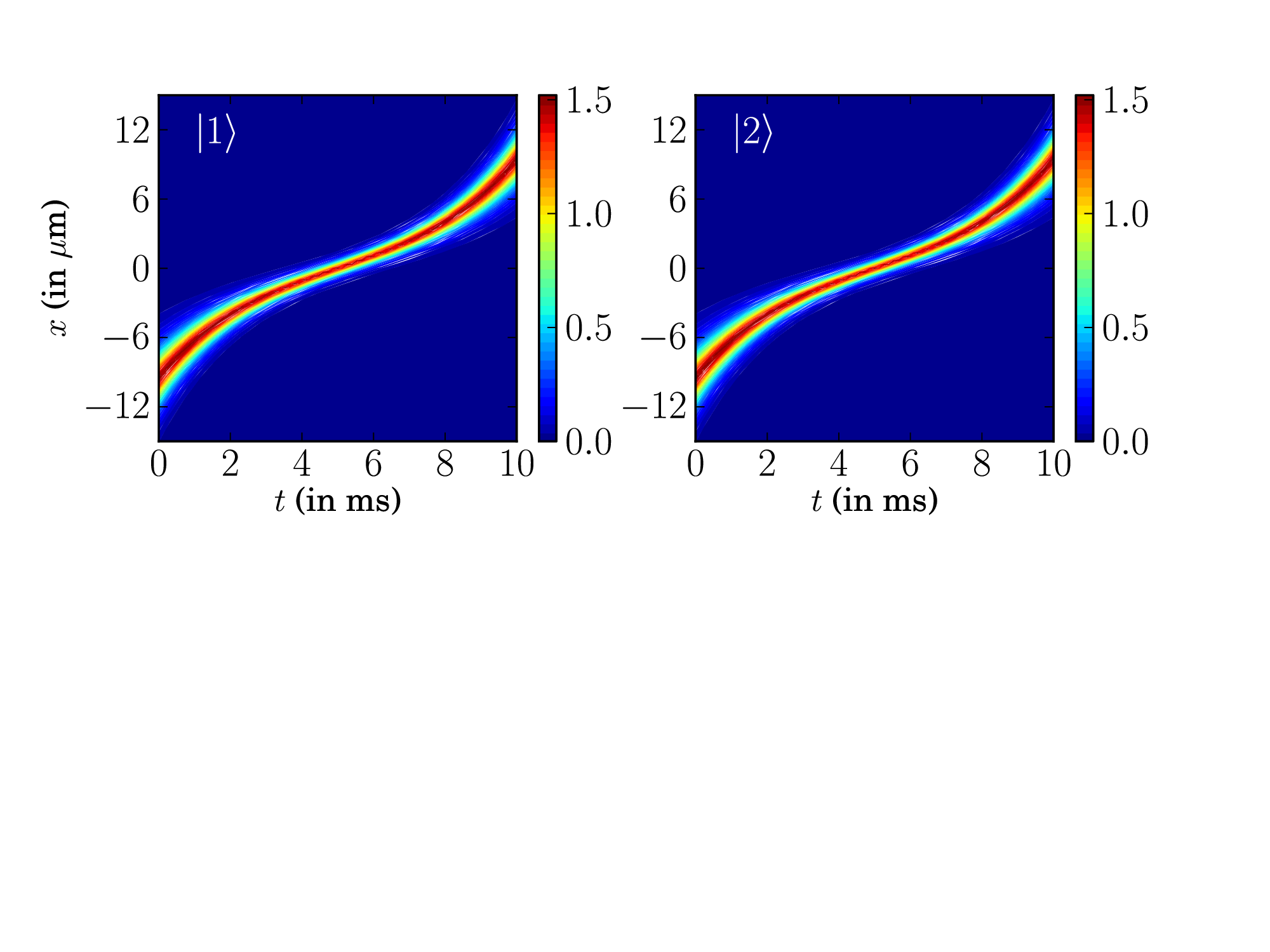}
\caption{(color online) One-soliton solution for time-independent expulsive harmonic trap potential $V(x) =-\Omega_0^2 x^2/2$ and $\gamma(t) = \Omega_0 \tanh(\Omega_0 t + \delta)$.  The parameters are $k_1=1+0.5i$, $\alpha_1^{(1)}=\alpha_1^{(2)}=1.0$, $\Omega_0=890/53$ and $r_0=0.5$.}
\label{fig:1sol_1}
\end{center}
\end{figure}

For $N=1$, we get the one-soliton solution of the coupled GP equation~(\ref{cgpe}) from Eq.~(\ref{sol:cgpe}).  As noted above, if we choose $\gamma(t) =  \Omega_0 \tanh(\Omega_0 t + \delta)$, the intensity of the soliton is constant.  Figure \ref{fig:1sol_1} shows the bright-bright one-soliton solution for the two components $ \vert 1 \rangle  = \vert \Psi_1 \vert$  and $ \vert 2 \rangle  = \vert \Psi_2 \vert$ for the above gain term where the amplitude of the wave packet remains constant.
For the $N=2$ case we get the two-soliton solution of ~(\ref{cgpe}) from Eq.~(\ref{sol:cgpe}). In this case, elastic collision occurs only for the specific choice of the parameters $(\alpha_1^{(1)}/\alpha_2^{(1)}) = (\alpha_1^{(2)}/\alpha_2^{(2)})$, see Eqs.~(\ref{eqn.20}) and (\ref{sol:cgpe}). For all other choices of the parameter values, shape changing/matter redistribution interaction occurs.
\begin{figure}[!ht]
\begin{center}
\includegraphics[width=\linewidth,clip]{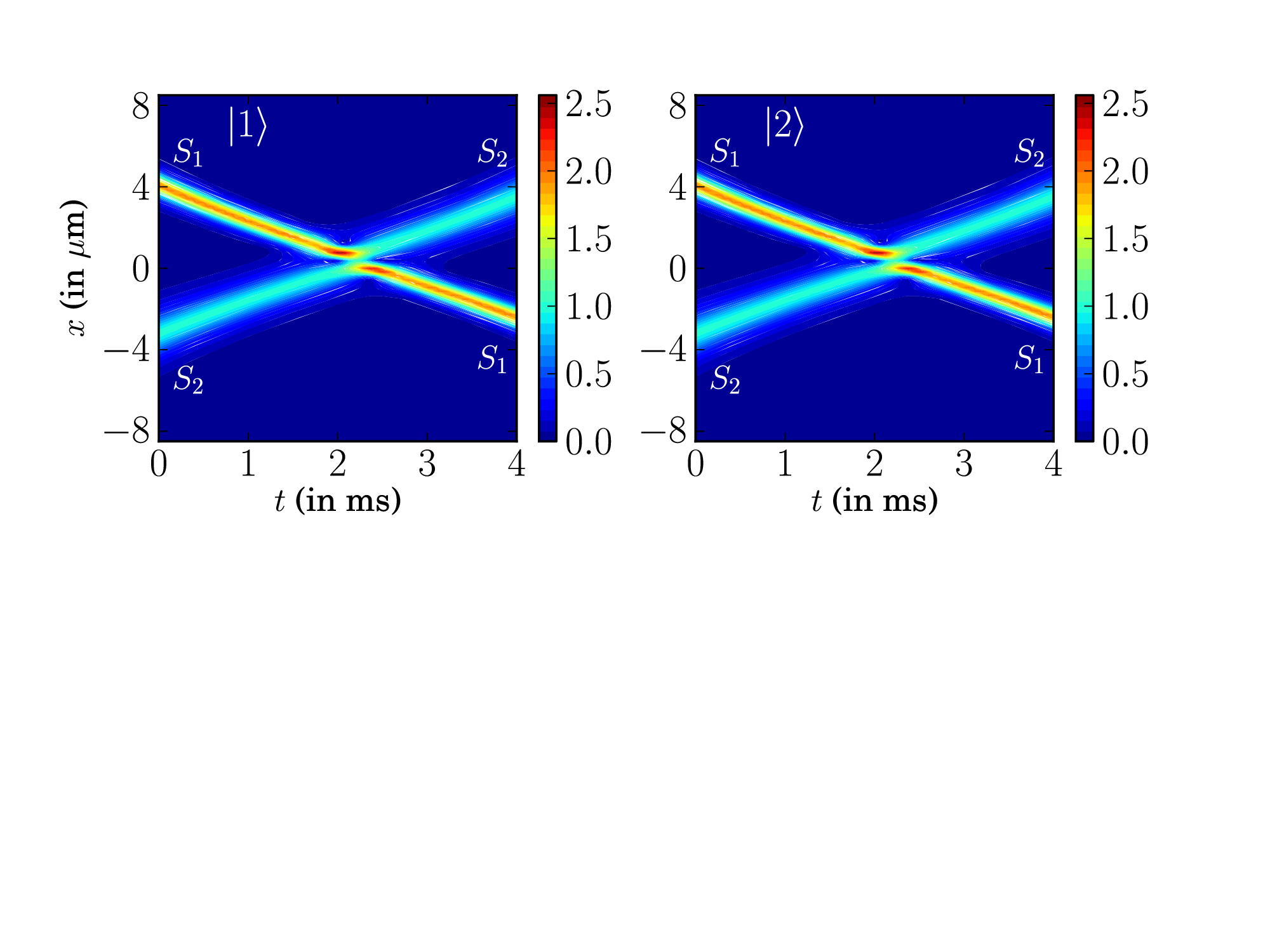}
\caption{(color online) Elastic interaction of bright-bright two solitons  for time-independent expulsive harmonic trap potential $V(x) = -\Omega_0^2 x^2/2$ and $\gamma(t) = \Omega_0 \tanh(\Omega_0 t + \delta)$ for $\alpha_1^{(1)}=\alpha_1^{(2)}=\alpha_2^{(1)}=\alpha_2^{(2)}=1$. The parameters are $k_1=1+i$, $k_2=2-i$, $\Omega_0=0.06$ and $r_0=0.5$. }
\label{fig:2sol_1e}
\end{center}
\end{figure}
\begin{figure}
\begin{center}
\includegraphics[width=\linewidth,clip]{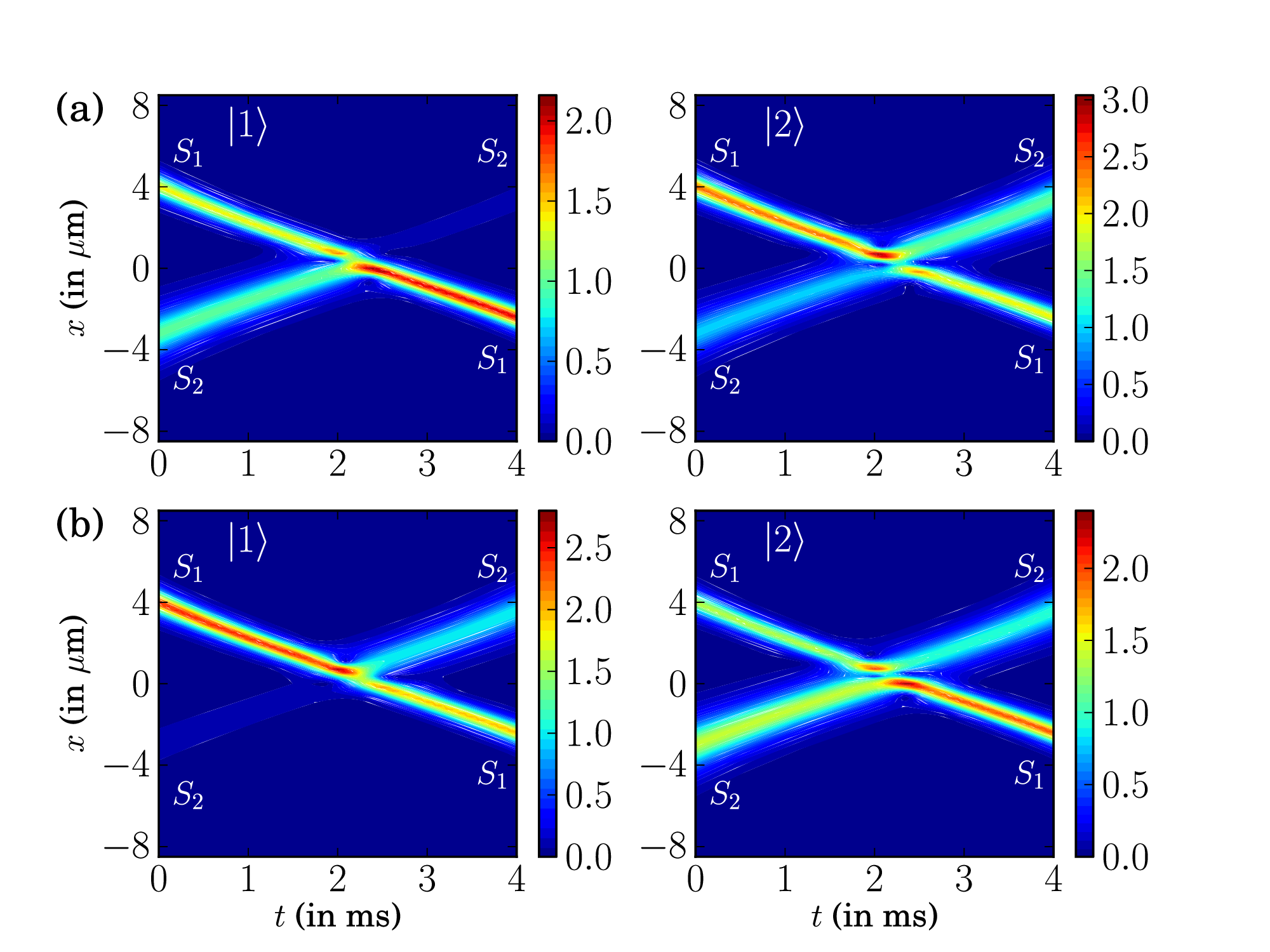}
\caption{(color online) Shape changing interactions of bright-bright two solitons for time-independent expulsive harmonic trap potential $V(x) = -\Omega_0^2 x^2/2$ and $\gamma(t) = \Omega_0 \tanh(\Omega_0 t + \delta)$ for  (a): $\alpha_1^{(1)}=\alpha_1^{(2)}=\alpha_2^{(2)}=1$, $\alpha_2^{(1)}=(39+80i)/89$ and for (b): $\alpha_1^{(1)}=(0.02+0.1i)$, $\alpha_2^{(1)}=\alpha_1^{(2)}=\alpha_2^{(2)}=1$. The other parameters are $k_1=1+i$, $k_2=2-i$, $\Omega_0=0.06$ and $r_0=0.5$. }
\label{fig:2sol_1sc}
\end{center}
\end{figure}
Fig.~\ref{fig:2sol_1e} shows the  elastic interaction of the bright-bright two-soliton solution for $k_1=1+i$, $k_2=2-i$, $\alpha_1^{(1)}=\alpha_1^{(2)}=\alpha_2^{(1)}=\alpha_2^{(2)}=1$. Here the intensity of the two solitons ($S_1$ and $S_2$) in both the components are unchanged before and after interaction. The two distinct possibilities of the shape changing two-soliton interaction are shown in Figs.~\ref{fig:2sol_1sc}(a) and \ref{fig:2sol_1sc}(b), see also Table~\ref{table:1}(a). Figs.~\ref{fig:2sol_1sc}(a) illustrate the shape changing two-soliton interaction for $k_1=1+i$, $k_2=2-i$, $\alpha_1^{(1)}=\alpha_1^{(2)}=\alpha_2^{(2)}=1$, $\alpha_2^{(1)}=(39+80i)/89$. Here the intensity of the soliton $S_1$ gets enhanced 
\begin{figure}[!ht]
\begin{center}
\includegraphics[width=\linewidth,clip]{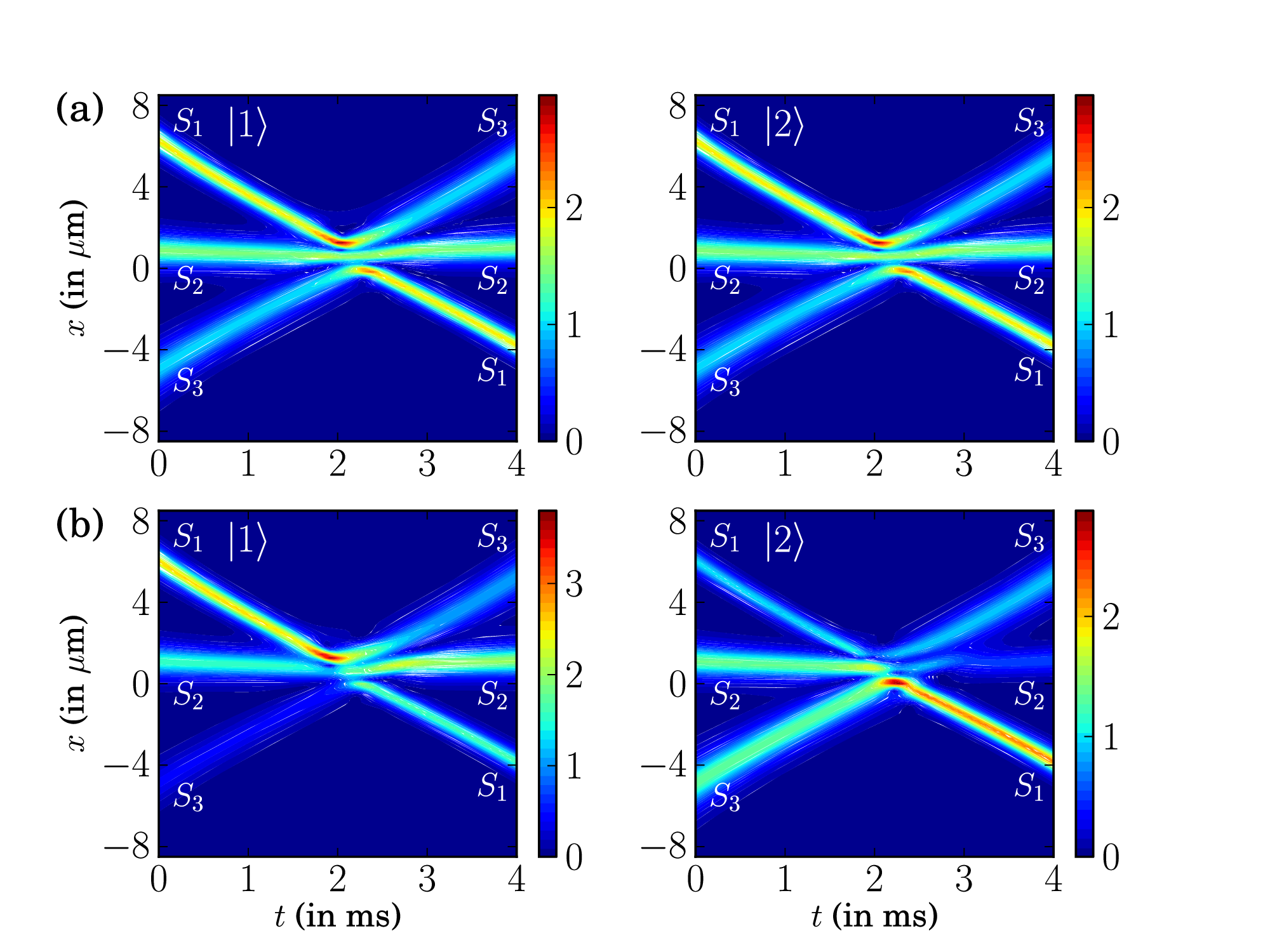}
\caption{(color online) Bright-bright three soliton interactions: (a) Elastic interaction for $\alpha_1^{(1)}=\alpha_1^{(2)}=\alpha_2^{(1)}=\alpha_2^{(2)}=\alpha_3^{(1)}=\alpha_3^{(2)}=1$ and (b) Shape changing  interaction  for $\alpha_1^{(1)}=0.39+0.1i$, $\alpha_2^{(1)}=0.3+0.2i$, $\alpha_3^{(1)}=1.0$, $\alpha_1^{(2)}=(80.0- 80 i)/89.0$, $\alpha_2^{(2)}=0.6+0.2$, $\alpha_3^{(2)}=(89.0+ 80 i)/89.0$  for time-independent expulsive harmonic trap potential $V(x) = -\Omega_0^2 x^2/2$ and $\gamma(t) = \Omega_0 \tanh(\Omega_0 t + \delta)$. The other parameters are $k_1=1+0.5i$, $k_2=1.5$, $k_3=2-1.5i$, $\Omega_0=0.06$ and $r_0=0.5$.}
\label{fig:3sol_1}
\end{center}
\end{figure}
while that of soliton $S_2$ is  suppressed after interaction in the component $\vert 1 \rangle$, whereas in the component $\vert 2 \rangle$ it gets reversed. Fig.~\ref{fig:2sol_1sc}(b) shows another possible  way of the shape changing two-soliton interaction for $k_1=1+i$, $k_2=2-i$,  $\alpha_1^{(1)}=(0.02+0.1i)$, $\alpha_2^{(1)}=\alpha_1^{(2)}=\alpha_2^{(2)}=1$. Here in contrast to the above  [cf. Figs.~\ref{fig:2sol_1sc} (a)],  the intensity of the soliton $S_1$ gets  suppressed while that of soliton $S_2$ is enhanced after interaction in the component $\vert 1 \rangle$, whereas in the component $\vert 2 \rangle$ it gets reversed similar to the well studied case of Manakov system~\cite{Radhakrishnan1997,kanna01,kanna03}.

For the $N=3$ case, we get the three-soliton solutions of the coupled GP equation~(\ref{cgpe}) from Eq.~(\ref{sol:cgpe}). Here elastic interaction occurs only for  $\alpha_1^{(1)}:\alpha_2^{(1)}:\alpha_3^{(1)} = \alpha_1^{(2)}:\alpha_2^{(2)}:\alpha_3^{(2)}$ and for all other choice of parameters, matter redistribution interaction occurs. Fig.~\ref{fig:3sol_1}(a) shows the elastic interaction of bright-bright $3$-soliton ($N=3$ case) solution for $k_1=1+0.5i$, $k_2=1.5$, $k_3=2-1.5i$, $\alpha_1^{(1)}=\alpha_1^{(2)}=\alpha_2^{(1)}=\alpha_2^{(2)}=\alpha_3^{(1)}=\alpha_3^{(2)}=1$. Here the intensities of the three solitons in  both the components are unchanged before and after interactions. Fig.~\ref{fig:3sol_1}(b) shows the shape changing interactions of bright-bright 3-soliton solution for $\alpha_1^{(1)}=0.39+0.1i$, $\alpha_2^{(1)}=0.3+0.2i$, $\alpha_3^{(1)}=1.0$, $\alpha_1^{(2)}=(80.0- 80 i)/89.0$, $\alpha_2^{(2)}=0.6+0.2$, $\alpha_3^{(2)}=(89.0+ 80 i)/89.0$. Here the intensity of the soliton $S_1$ gets suppressed  ($S$) while that of solitons $S_2$ and $S_3$ are  enhanced ($E$) after interaction in the component $\vert 1 \rangle$, whereas in the component $\vert 2 \rangle$ it gets reversed. The six distinct possibilities of shape changing interactions of three soliton solutions are shown in Table~\ref{table:1}(b).
\begin{table}[!ht]
\caption{Possible combinations of matter redistributions of two solitons ($N=2$ case) and three solitons ($N=3$ case) under collision in component $\vert 1 \rangle$. $E$ stands for enhancement of intensity while $S$ represent suppression. See also Ref.~\cite{kanna01}}
\label{table:1} \centering
\begin{tabular}{cccc}
  \hline \hline \multicolumn{1}{c}{}&\multicolumn{3}{c}{(a) $N=2$ case}\\
 case & \qquad $S_1$&\qquad $S_2$&
 \\
 \hline
 1 & \qquad $E$ & \qquad $S$ &
 \\
2 & \qquad $S$ & \qquad $E$ &
 \\
\hline&&& \\
\multicolumn{1}{c}{}&\multicolumn{3}{c}{(b) $N=3$ case}\\

 case & \quad $S_1$& \qquad $S_2$&\qquad $S_3$
 \\\hline
 1 & \quad $E$ &  \qquad $S$ &\qquad $S$
 \\
2 &\quad   $S$ &\qquad   $E$ &\qquad $S$
 \\
3 & \quad  $S$ & \qquad  $S$ & \qquad $E$
 \\
4 & \quad $S$ &  \qquad $E$ &\qquad $E$
 \\
 5 &\quad   $E$ & \qquad  $S$ &\qquad $E$
 \\
6 &\quad  $E$ & \qquad  $E$ &\qquad $S$
 \\
\hline \hline
\end{tabular}
\end{table}
The above three soliton interaction process is equivalent to two pairwise interactions. Kanna and Lakshmanan ~\cite{kanna01,kanna03} have shown that for the corresponding Manakov system the first interaction is controlled by the parameters $\alpha_1^{(1)}$, $\alpha_2^{(1)}$, $\alpha_1^{(2)}$, $\alpha_2^{(2)}$, $k_1$, $k_2$ and the second interaction is controlled by  $\alpha_3^{(1)}$, $\alpha_3^{(2)}$ and $k_3$.

\begin{figure}[!ht]
\begin{center}
\includegraphics[width=\linewidth,clip]{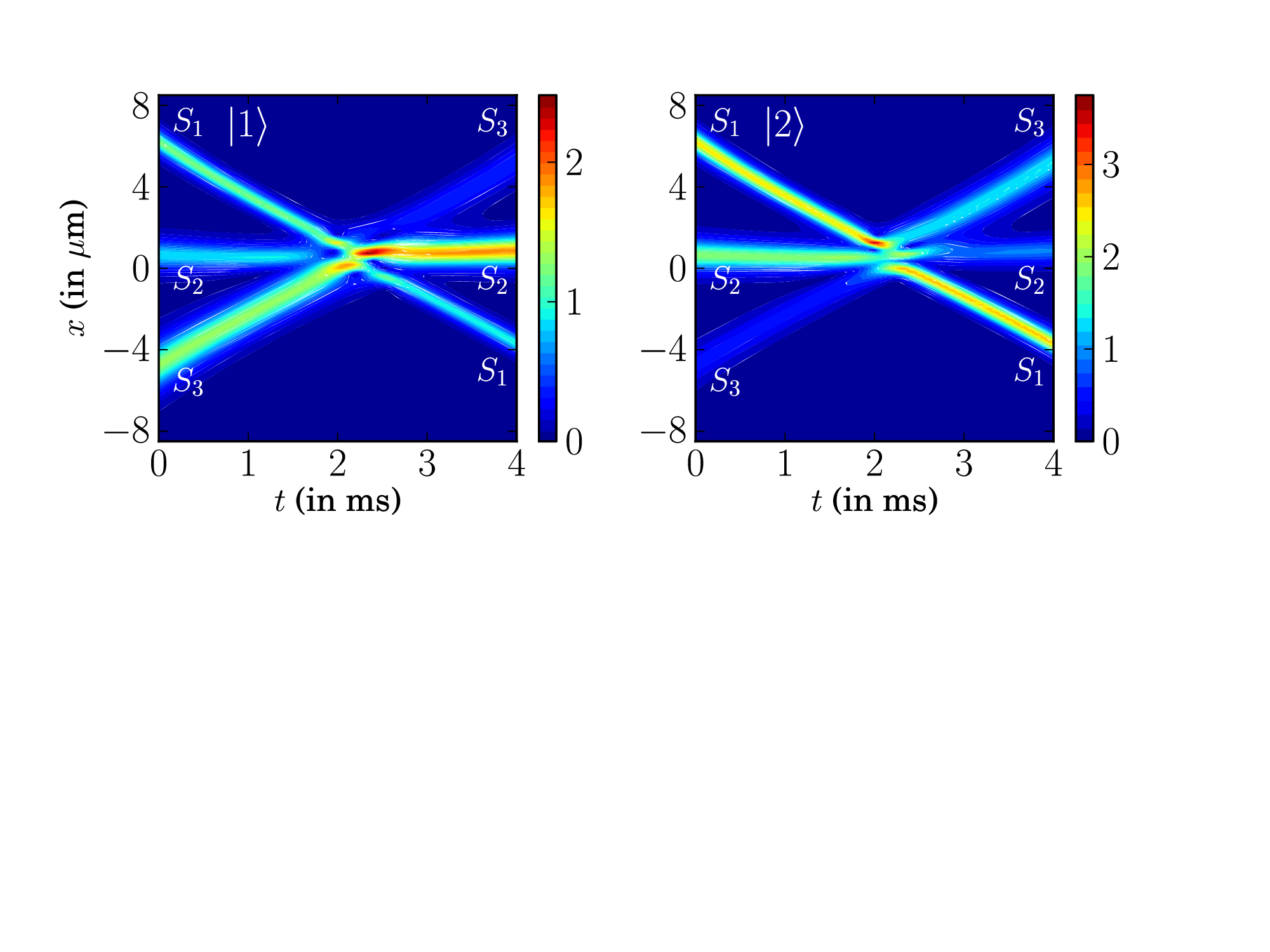}
\caption{(color online) Shape restoring property of $S_1$ in three solitons interactions for time-independent expulsive harmonic trap potential $V(x) = -\Omega_0^2 x^2/2$ and $\gamma(t) = \Omega_0 \tanh(\Omega_0 t + \delta)$ for the choice of the parameters $\alpha_1^{(1)}=(39-80i)/89$, $\alpha_2^{(1)}=(39+80i)/89$, $\alpha_3^{(1)}=0.3+0.2i$, $\alpha_1^{(2)}=0.39$, $\alpha_2^{(2)}=\alpha_3^{(2)}=1$, $k_1=1+1.5i$, $k_2=1.5$, $k_3=2-1.5i$, $\Omega_0=0.06$ and $r_0=0.5$.}
\label{fig:3sol_1sr}
\end{center}
\end{figure}

Fig.~\ref{fig:3sol_1sr} depicts the shape restoring property of soliton $S_1$ during its interaction with the other two solitons,  $S_2$ and $S_3$ for time-independent 
expulsive harmonic trap potential $V(x) = -\Omega_0^2 x^2/2$ for the choice of the parameters $\alpha_1^{(1)}=(39-80i)/89$, $\alpha_2^{(1)}=(39+80i)/89$, $\alpha_3^{(1)}=0.3+0.2i$, $\alpha_1^{(2)}=0.39$, $\alpha_2^{(2)}=\alpha_3^{(2)}=1$, $k_1=1+1.5i$, $k_2=1.5$, $k_3=2-1.5i$. Here the intensity of the soliton $S_1$ is unchanged after its interactions with the other two solitons $S_2$ and $S_3$ in both the components, while the intensities of solitons $S_2$ and $S_3$ get changed. The condition for the choice of the parameters for the shape restoration of soliton $S_1$ is given in~Ref.~\cite{kanna03}. Note that the shape restoring property is essential for the construction of universal logic gates which are necessary for computing~\cite{Steiglitz2000,kanna03}. The above type of elastic, shape changing interactions and  shape restoring property of solitons during the interactions are similar to the study of Kanna and Lakshmanan~\cite{kanna01,kanna03} in the context of optical computing, where the intensities of light pulses are transformable between two modes. Here, the shape changing interactions are interpreted as the transform of the fraction of atoms between the two components, which can be achieved experimentally by suitable tuning of atomic scattering length and gain/loss term. Note that exact analytical representations for the soliton switching can be given in the form of linear fractional tranformations leading to logic gates as in the case of optical systems, see Ref.~\cite{Steiglitz2000,kanna03}. This type of shape changing soliton interactions can be used (as disscussed in Refs.~\cite{Folman2000,Folman2001,Petrov2009}) in the matter wave switching devices, logic gates and quantum information processing as in the case of optical computing.

\subsection{Periodically Modulated Trap Potential}

\begin{figure}[!ht]
\begin{center}
\includegraphics[width=0.9\linewidth,clip]{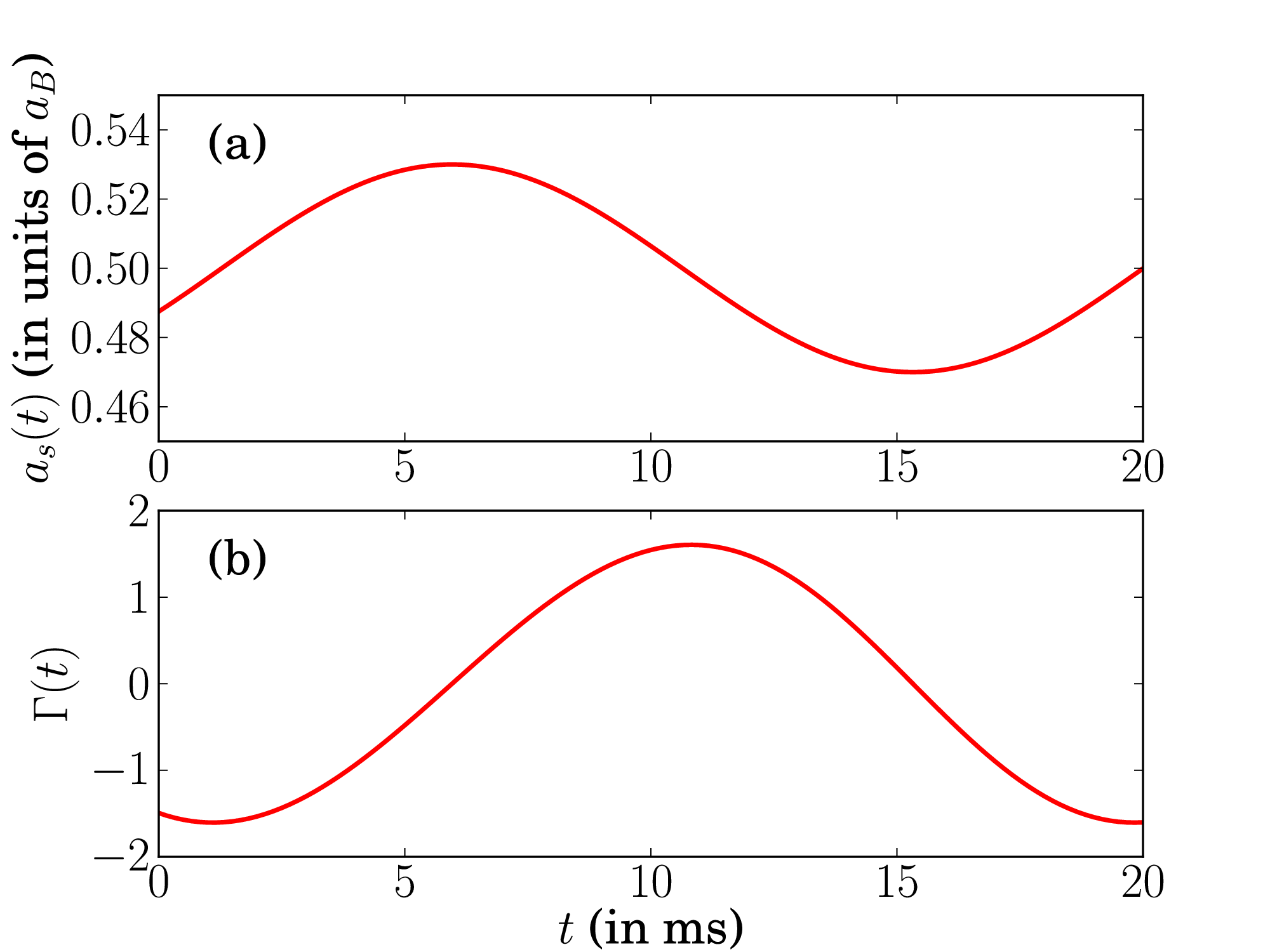}
\caption{(color online) Form of the (a) atomic scattering length $a_s(t)$ and (b) gain or loss term $\Gamma(t)$ as a function of time for the periodically modulated harmonic potential with the trap frequency~(\ref{poten_pm}).}
\label{fig_gamma_two}
\end{center}
\end{figure}

 Next, if we choose  the periodically varying atomic scattering length~\cite{Baizakov2005} $R(t) =1+ \omega \cos(\omega t+\delta)$  and the corresponding gain term
\begin{align}
\gamma(t)=\displaystyle \frac{ \omega^2\sin(\omega t+ \delta)}{2\left[1+  \omega \cos(\omega t+\delta)\right]}, \;\;\; \omega \le 1,\label{gamma_2}
\end{align}
with $\delta$ as a constant, we get $\tilde{R}(t) = \sqrt{1+ \omega \cos(\omega t+\delta)}$ from Eq.~(\ref{rtilde})and the periodically modulated trap frequency from the integrability condition (\ref{riccati}) as
\begin{align}
\Omega^2(t)= \omega^2\left\{1-\frac{4+ 10  \omega \cos \tilde{\theta}+ 3 \omega^2 \left[1+\cos^2\tilde{\theta}\right]}{4 \left[1+ \omega\cos\tilde{\theta}\right]^2 }\right\},\label{poten_pm}
\end{align}
where $\tilde{\theta} = \omega t+\delta $. The intensity of soliton also remains constant for this case  but the width of the soliton is periodically varying with time (that is, inversely proportional to $\sqrt{\tilde{R}(t)}$). Fig.~\ref{fig_gamma_two}  shows the gain term
 \begin{align}\Gamma(t)=  \displaystyle  \frac{ \omega_\perp \omega^2 \sin(\omega t+ \delta)}{2[1+  \omega\cos(\omega t+\delta)]},
\end{align}
which can can also be experimentally realized by suitable optical pumping, and the corresponding choice of atomic scattering length $a_s(t)=\frac{a_B}{2}\left[1+ \omega\cos(\omega t+\delta)\right] $
\begin{figure}[!ht]
\begin{center}
\includegraphics[width=\linewidth,clip]{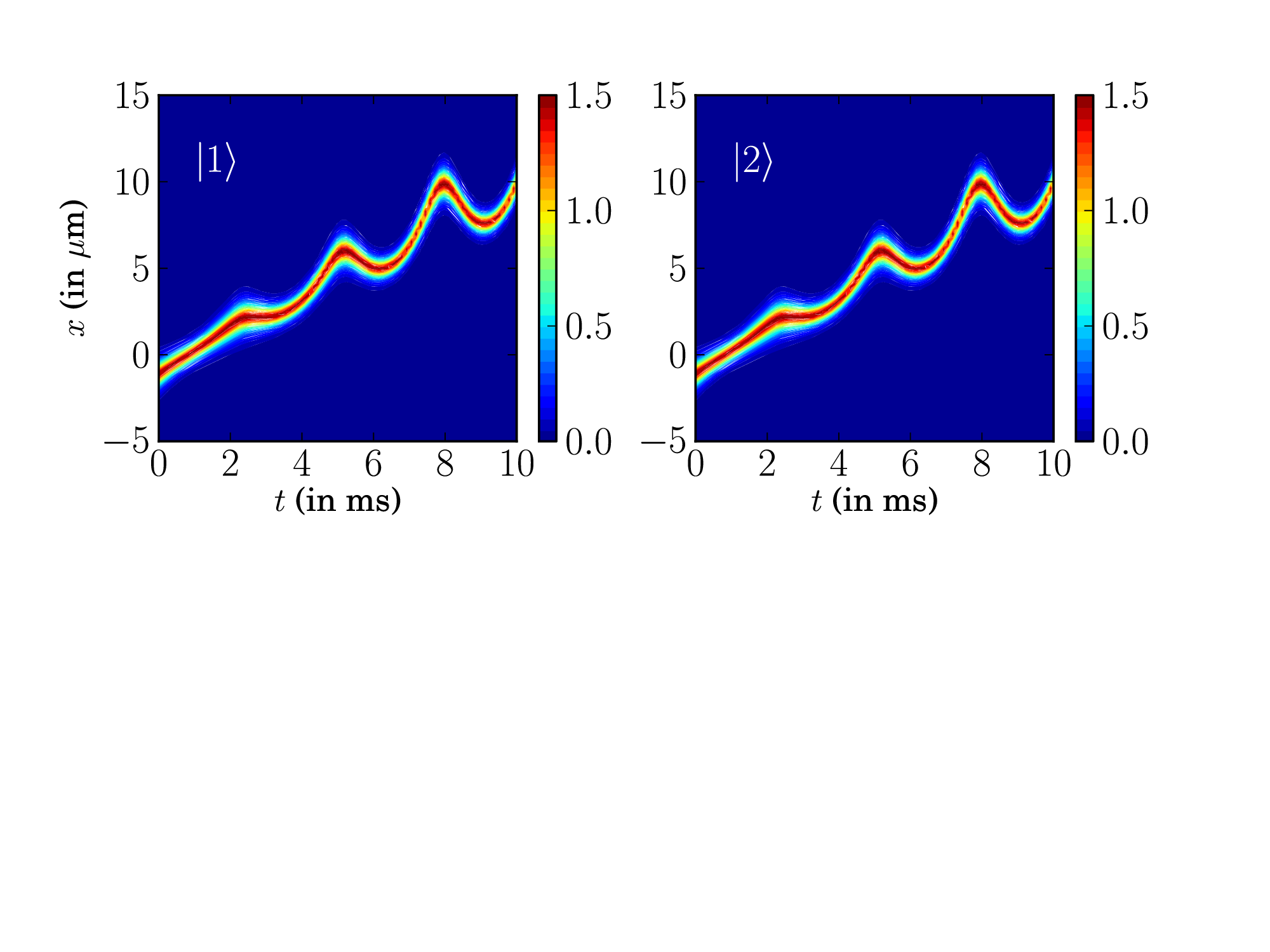}
\caption{(color online) One-soliton solution for the periodically modulated harmonic potential with trap frequency~(\ref{poten_pm}). The parameters are  fixed as $k_1=1+0.5i$, $\alpha_1^{(1)}=\alpha_1^{(2)}=1.0$, $\omega=0.4$ and $r_0=0.5$.}
\label{fig:1sol_2}
\end{center}
\end{figure}
which may be experimentally realized by periodically tuning the external magnetic field as
\begin{align}
B(t)=B_0+ \frac{a_s^0 \,\,\,\Delta}{a_s^0-\frac{1}{2} a_B \left[1+ \omega\cos(\omega t+\delta)\right] }.
\end{align}
\begin{figure}[!ht]
\begin{center}
\includegraphics[width=\linewidth,clip]{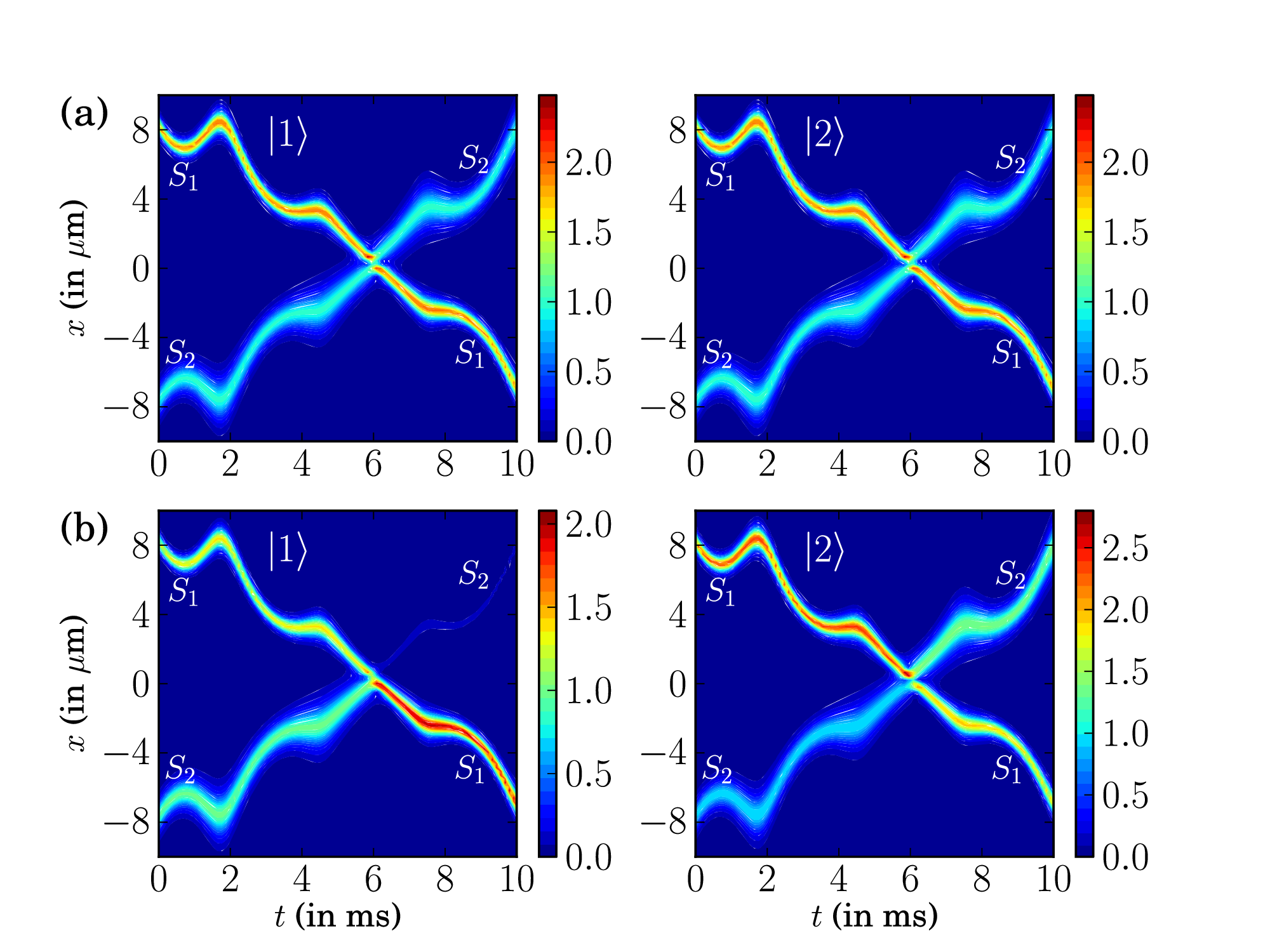}
\caption{(color online) Bright-bright two soliton interactions for the periodically modulated harmonic potential. (a) Elastic collision for $\alpha_1^{(1)}=\alpha_1^{(2)}=\alpha_2^{(1)}=\alpha_2^{(2)}=1$ and (b) Shape changing collision for $\alpha_1^{(1)}=\alpha_1^{(2)}=\alpha_2^{(2)}=1$, $\alpha_2^{(1)}=(39+80i)/89$. The other parameters are fixed as $k_1=1+i$, $k_2=2-i$, $\omega=0.4$ and $r_0=0.5$. }
\label{fig:2sol_2}
\end{center}
\end{figure}
\begin{figure}[!ht]
\begin{center}
\includegraphics[width=\linewidth,clip]{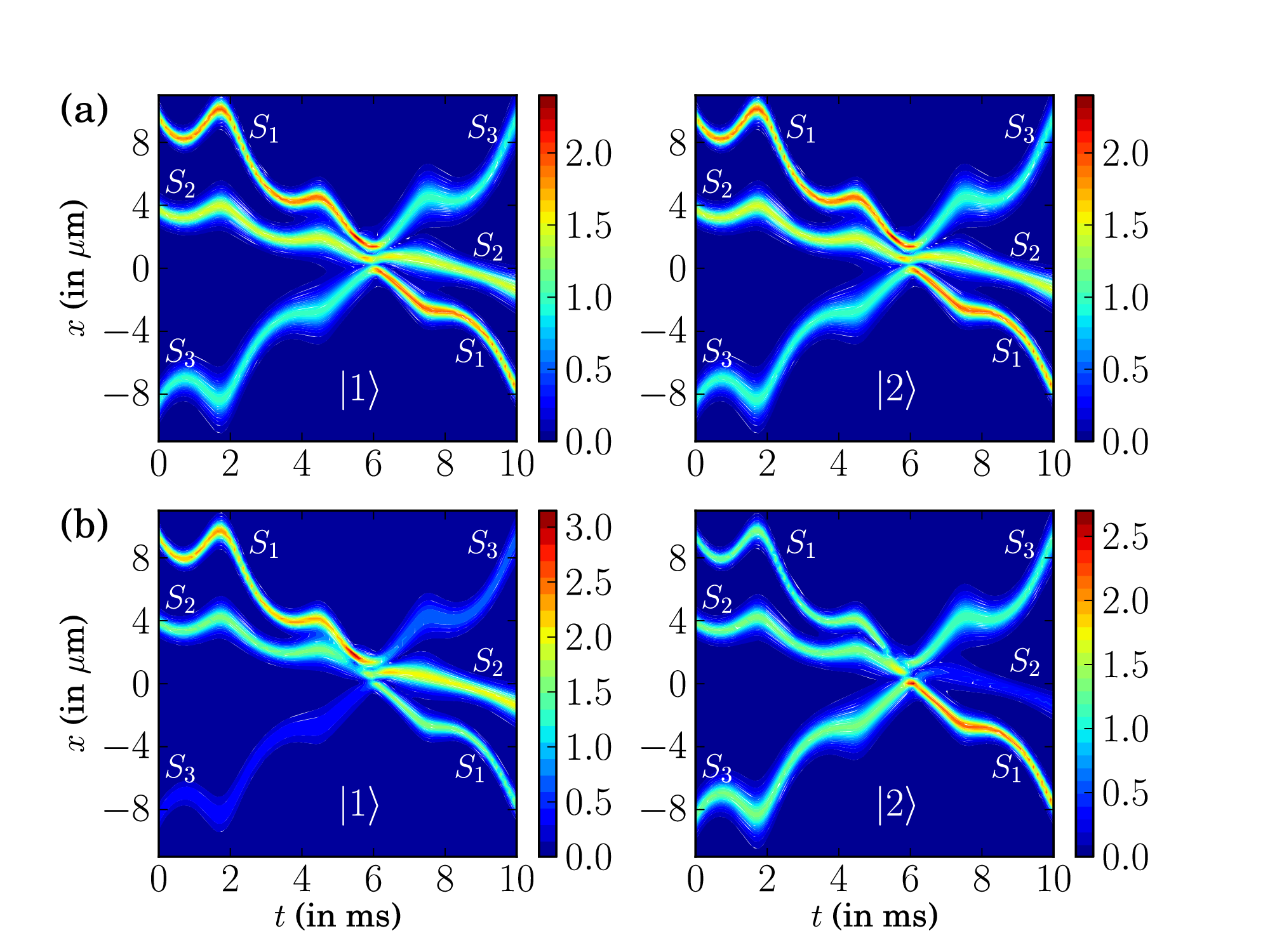}
\caption{(color online) Bright-bright three soliton interactions for the periodically modulated harmonic trap potential. (a) Elastic collision for $\alpha_1^{(1)}=\alpha_1^{(2)}=\alpha_2^{(1)}=\alpha_2^{(2)}=\alpha_3^{(1)}=\alpha_3^{(2)}=1$ and (b) Shape changing collision for $\alpha_1^{(1)}=0.39+0.1i$, $\alpha_2^{(1)}=0.3+0.2i$, $\alpha_3^{(1)}=1.0$, $\alpha_1^{(2)}=(80.0- 80 i)/89.0$, $\alpha_2^{(2)}=0.6+0.2$, $\alpha_3^{(2)}=(89.0+ 80 i)/89.0$. The parameters are $k_1=1+i$, $k_2=1.5-0.5i$, $k_3=2-i$, $\omega=0.4$ and $r_0=0.5$.}
\label{fig:3sol_2}
\end{center}
\end{figure}
Fig.~\ref{fig:1sol_2} shows the snake-like effect of the one-soliton solution for the above periodically modulated trap potential with $R(t) = \left[1+ \omega\cos(\omega t+\delta)\right] $ and
$\gamma(t)$ as given in Eq.~(\ref{gamma_2}),
where the intensity of the soliton remains constant in both the components while the width of the soliton varies periodically with time. Note that the oscillation of the soliton goes on increasing with time due to the fact that the width of the soliton is inversely proportional to $\tilde{R}(t)$. It is of importance to note that recently, in scalar nonautonomous NLS equation for BEC~\cite{Rajendran2010}  and  optical solitons~\cite{serkin2000}, a similar kind of snake-like effect has been demonstrated.

Next we analyze different types of two soliton interactions for the periodically modulated potential for suitable choice of other parameters. Fig.~\ref{fig:2sol_2}(a) shows the  elastic collision of snake like bright-bright two solitons while Fig.~\ref{fig:2sol_2}(b) shows shape changing collision of snake-like bright-bright two solitons for $\omega=0.4$. Figs.~\ref{fig:3sol_2}(a) and (b) depict the elastic collision and shape changing collision of snake-like bright-bright three-soliton solutions, respectively, for this case.  The other parameters are the same as in the time independent expulsive harmonic potential case. The collision effects are similar to the ones discussed in the case of expulsive trap potential earlier but now the widths of solitons oscillate periodically with time because of the periodically varying atomic scattering length.

\subsection{Kink-like Modulated Trap Potential}

Finally, if we choose  $R(t) =1+\tanh(\omega t+\delta)$ and
\begin{align}
\gamma(t)=  \displaystyle -\frac{\omega}{4} \left[1-\tanh(\omega t+\delta)\right],
\end{align}
  we get $\tilde{R}(t) = \sqrt{1+\tanh(\omega t+\delta)}$ and the integrability condition (\ref{riccati}) gives  \begin{align}\Omega^2(t)= -\omega^2 \left[1-\frac{\mbox{e}^{4 \omega t}}{(1+\mbox{e}^{2 \omega t})^2}\right],\end{align} which is a kink-like modulated trap.
\begin{figure}[!ht]
\begin{center}
\includegraphics[width=0.9\linewidth,clip]{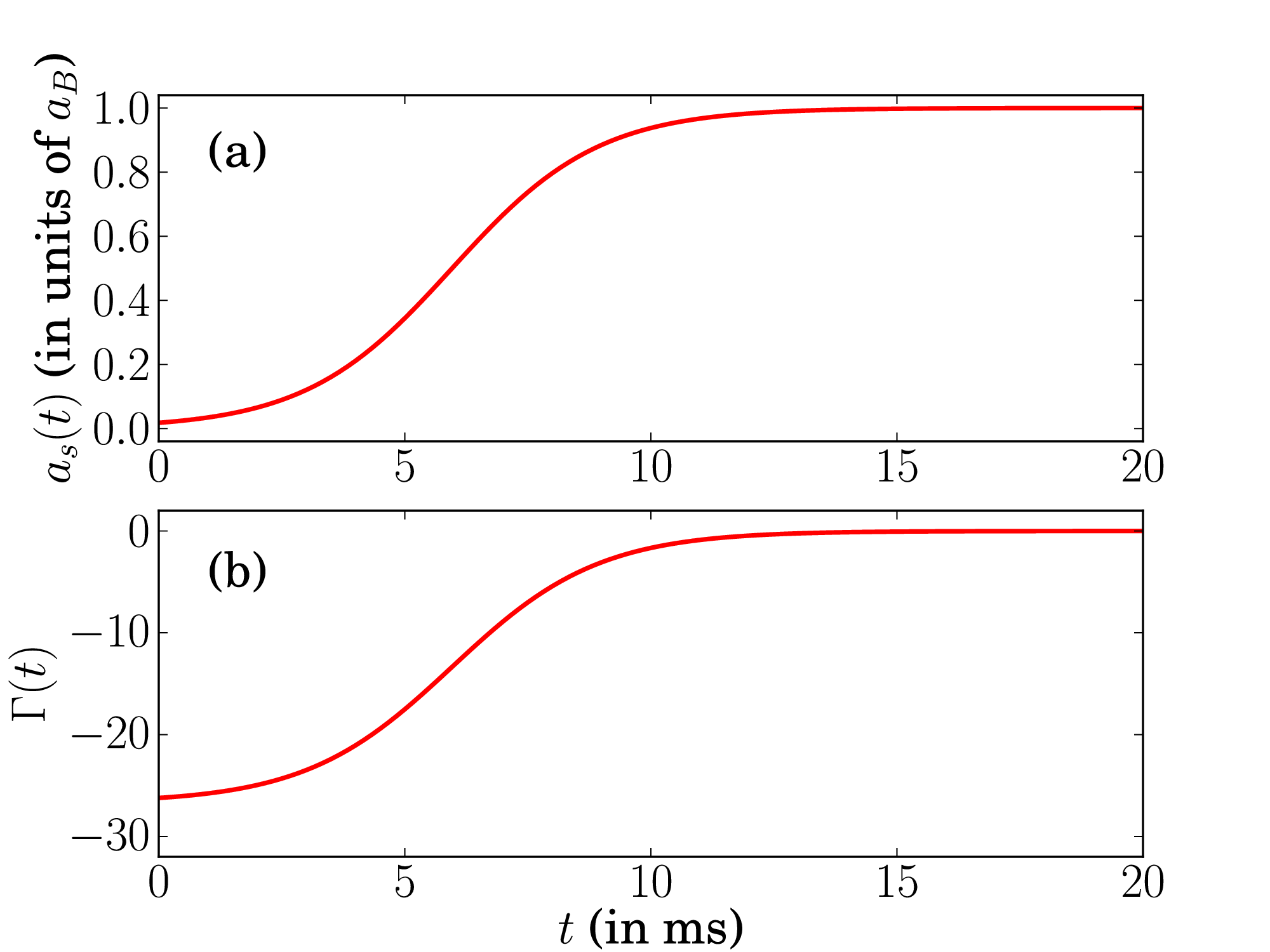}
\caption{(color online) Choice of the (a) atomic scattering length $a_s(t)$ and (b) gain term $\Gamma(t)$ as a function of time for the kink-like modulated harmonic potential $V(x,t) = \displaystyle-\omega^2 \left(1-\frac{\mbox{e}^{4 \omega t}}{(1+\mbox{e}^{2 \omega t})^2}\right) x^2$.}
\label{fig_gamma_three}
\end{center}
\end{figure}
For the above case, we sketch the gain 
\begin{align}
\Gamma(t) = \displaystyle -\frac{\omega_{\perp}\omega}{4} \left[1-\tanh(\omega t+\delta)\right],\label{gamma_3}
\end{align}
 and the corresponding
\begin{figure}[!ht]
\begin{center}
\includegraphics[width=\linewidth,clip]{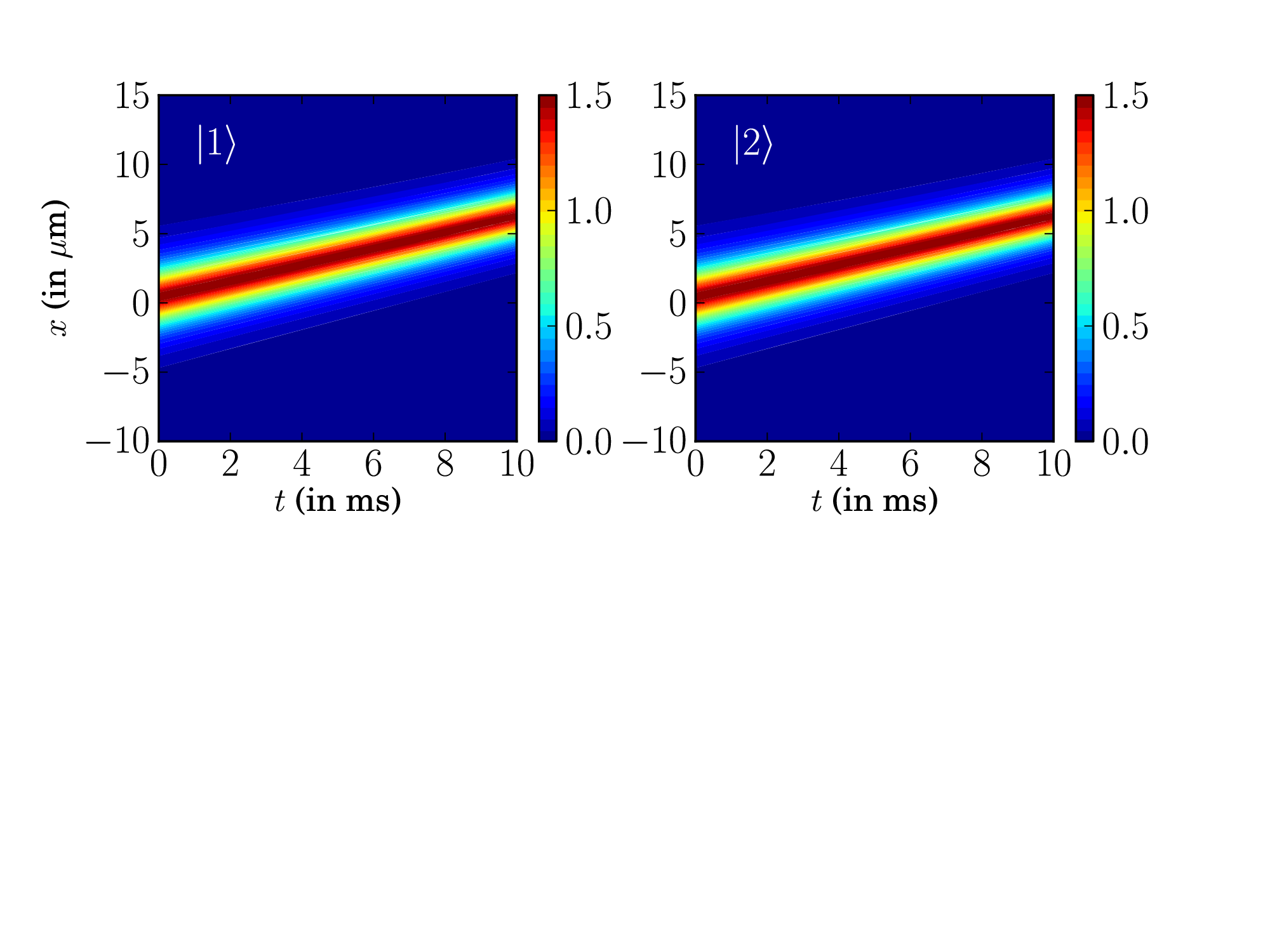}
 \caption{(color online) One-soliton solution for the kink-like trap potential. The parameters are $k_1=1+0.5i$, $\alpha_1^{(1)}=\alpha_1^{(2)}=1.0$, $\omega=0.06$ and $r_0=0.75$.}
\label{fig:1sol_3}
\end{center}
\end{figure}
\begin{figure}[!ht]
\begin{center}
\includegraphics[width=\linewidth,clip]{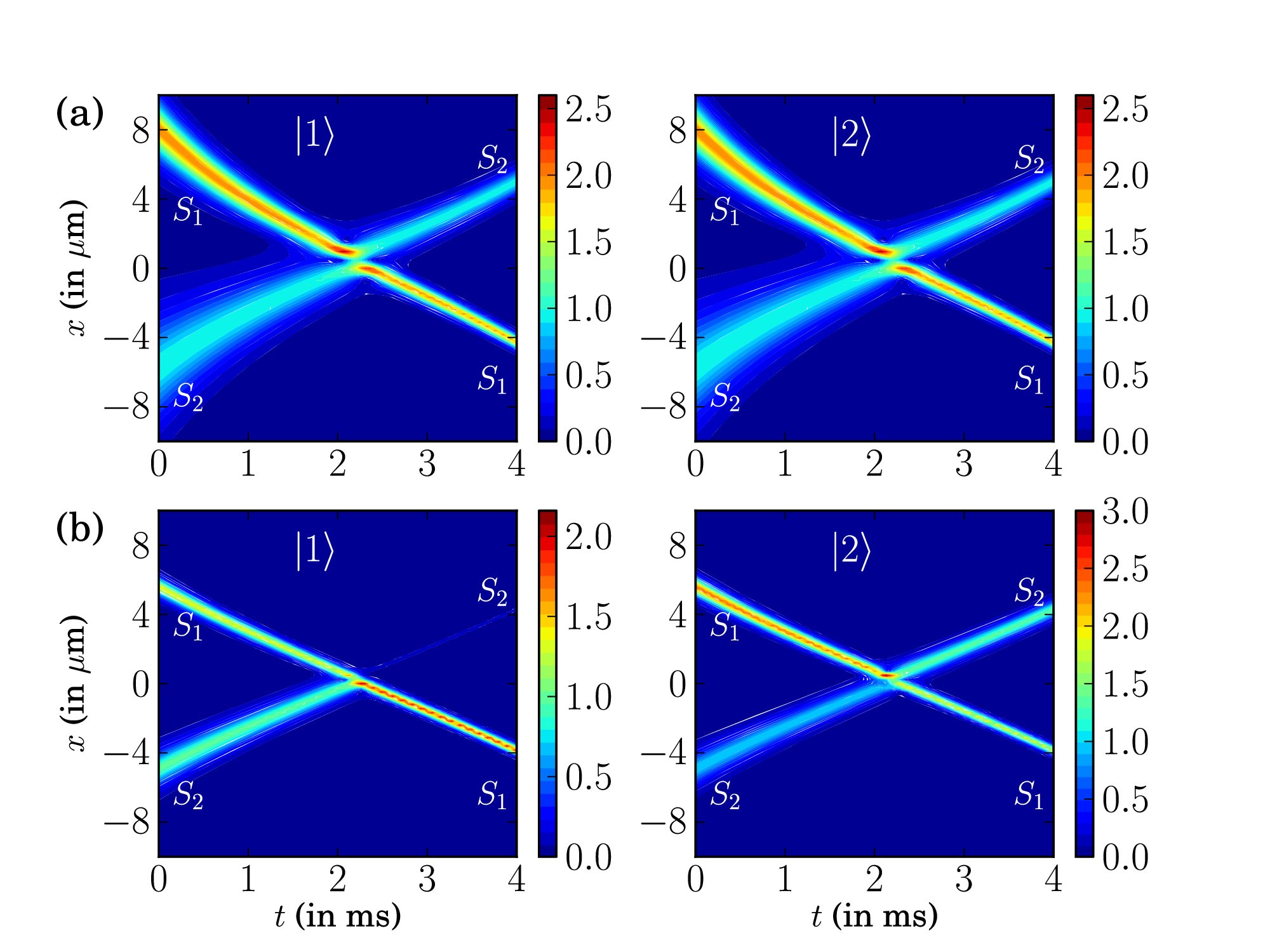}
\caption{(color online) Bright-bright two soliton interactions for the kink-like modulated harmonic trap potential. (a) Intensity unchanged under the collision for $\alpha_1^{(1)}=\alpha_1^{(2)}=\alpha_2^{(1)}=\alpha_2^{(2)}=1$. (b) Matter redistribution collision for $\alpha_1^{(1)}=\alpha_1^{(2)}=\alpha_2^{(2)}=1$, $\alpha_2^{(1)}=(39+80i)/89$. The parameters are $k_1=1+i$, $k_2=2-i$, $\omega=0.06$ and $r_0=0.4$ }
\label{fig:2sol_3}
\end{center}
\end{figure}
choice of atomic scattering length $a_s(t)=\frac{a_B}{2}\left[1+\tanh(\omega t+\delta)\right] $ in Fig.~\ref{fig_gamma_three}. The form~(\ref{gamma_3}) of the gain can again be experimentally realized by continuously loading the external atoms into the condensate by optical pumping as in the case of Refs.~\cite{Davis2000,kohl2002} and the atomic scattering length can be realized by a kink-like tuning of the external magnetic field as
 \begin{align}
B(t)=B_0+ \frac{a_s^0 \,\,\,\Delta}{a_s^0-\frac{1}{2} a_B \left[1+\tanh(\omega t+\delta)\right] }.
\end{align}
Fig.~\ref{fig:1sol_3} shows the one-soliton solution for the kink-like modulated trap potential with $R(t) =  1+\tanh(\omega t+\delta) $ and \begin{align}\gamma(t)=  \displaystyle -\frac{\omega}{4} \left[1-\tanh(\omega t+\delta)\right]\notag.\end{align}\begin{figure}[!ht]
\begin{center}
\includegraphics[width=\linewidth,clip]{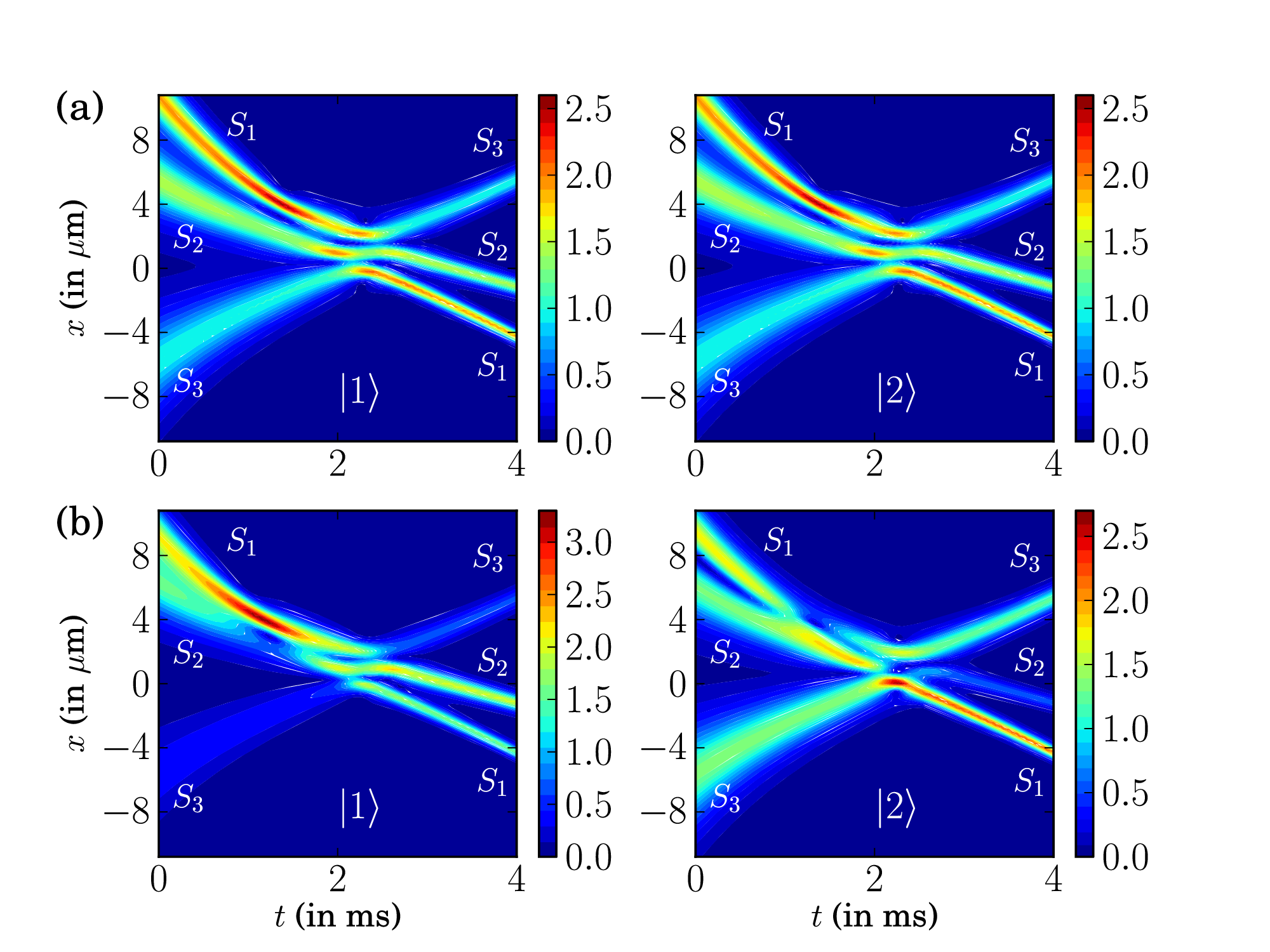}
\caption{(color online) Bright-bright three soliton interactions for the kink-like modulated harmonic trap potential. (a) Intensity unchanged  collision for $\alpha_1^{(1)}=\alpha_1^{(2)}=\alpha_2^{(1)}=\alpha_2^{(2)}=\alpha_3^{(1)}=\alpha_3^{(2)}=1$. (b) Matter redistribution collision for $\alpha_1^{(1)}=0.39+0.1i$, $\alpha_2^{(1)}=0.3+0.2i$, $\alpha_3^{(1)}=1.0$, $\alpha_1^{(2)}=(80.0- 80 i)/89.0$, $\alpha_2^{(2)}=0.6+0.2$, $\alpha_3^{(2)}=(89.0+ 80 i)/89.0$. The parameters are $k_1=1+i$,$k_2=1.5-0.5i$, $k_3=2-i$,  $\omega=0.06$ and $r_0=0.4$}
\label{fig:3sol_3}
\end{center}\end{figure}
Here the intensity of solitons in both the components are constant while the width is decreasing with time (that is inversely proportional to $\sqrt{\tilde{R}(t)}$).

Next we analyze the two-soliton ($N=2$) and three-soliton ($N=3$) solutions for the two-component BECs with kink-like modulated harmonic trap potential. Fig.~\ref{fig:2sol_3}(a) shows the bright-bright intensity unchanged collision of two solitons while Fig.~\ref{fig:2sol_3}(b) shows matter redistribution collision of two solitons. The intensity unchanged and matter redistribution collisions of three-soliton solutions for the kink-like modulated potential are shown in Figs.~\ref{fig:3sol_3}(a) and \ref{fig:3sol_3}(b), respectively. The other parameters are the same as in the time-independent potential case. Note that, here, the  widths of the two and three solitons are also decreasing with time.

\section{Summary and conclusion}\label{sec.Conclusion}

In summary, we have investigated the exact bright-bright multi-soliton solutions of the two-component BECs with time varying parameters such as trap frequency, s-wave scattering length and gain/loss term. On mapping the two coupled GP equations onto the coupled NLS equations under certain conditions, we have deduced different kinds of bright-bright one-soliton solutions and interaction of two solitons for time independent expulsive harmonic trapping potential, periodically modulated trap potential and kink-like modulated potential. We have shown the shape changing and elastic interactions of bright-bright two-soliton and three-soliton solutions of the two component BECs. The present study provides an understanding of the possible mechanism for the fraction of atoms transform between the two components. Especially the shape changing collisions of matter wave solitons can used for matter wave switching devices, logic gates and quantum information processing. These elastic and shape changing interactions can be realized in experiments by suitable control of time dependent trap parameters, atomic interaction and interaction with thermal cloud.

\acknowledgments

This work is supported by Department of Science and Technology (DST), Government of India - DST-IRHPA project (SR and ML), and DST Ramanna Fellowship (ML). The work of PM forms part of a DST project (Ref. No. SR/S2/HEP-003/2009).


\end{document}